\documentclass[useAMS,usenatbib]{mn2e}
\usepackage{natbib}
\usepackage{epsf}
\usepackage{graphicx}

\def\simlt{\mathrel{\rlap{\lower 3pt\hbox{$\sim$}}
        \raise 2.0pt\hbox{$<$}}}
\def\simgt{\mathrel{\rlap{\lower 3pt\hbox{$\sim$}}
        \raise 2.0pt\hbox{$>$}}}
\def\1550{XTE J1550--564}

\title[A jet dominating the spectrum of XTE J1550--564]{Evidence for a compact jet dominating the broadband spectrum of the black hole accretor XTE J1550--564}
\author[D. M. Russell et al.]{D. M. Russell$^{1}$\thanks{E-mail: d.m.russell@uva.nl}, D. Maitra$^{1,2}$, R. J. H. Dunn$^{3,4}$, S. Markoff$^{1}$\\
$^{1}$Astronomical Institute `Anton Pannekoek', University of Amsterdam, P.O. Box 94249, 1090 GE Amsterdam, the Netherlands\\
$^{2}$Department of Astronomy, University of Michigan, 500 Church Street, Ann Arbor, MI 48109, USA\\
$^{3}$Technische Universit\"at M\"unchen, Excellence Cluster Universe, Boltzmannstrasse 2, D-85748 Garching, Germany\\
$^{4}$Alexander von Humboldt Fellow\\
}
\begin{document}


\pagerange{\pageref{firstpage}--\pageref{lastpage}} \pubyear{2009}

\maketitle

\label{firstpage}

\begin{abstract}
The black hole X-ray binary \1550 was monitored extensively at X-ray, optical and infrared wavelengths throughout its outburst in 2000. We show that it is possible to separate the optical/near-infrared (OIR) jet emission from the OIR disc emission. Focussing on the jet component, we find that as the source fades in the X-ray hard state, the OIR jet emission has a spectral index consistent with optically thin synchrotron emission ($\alpha \approx -0.6$ to $-0.7$, where $F_{\nu} \propto \nu^{\alpha}$). This jet emission is tightly and linearly correlated with the X-ray flux; $L_{\rm OIR,jet} \propto L_{\rm X}^{0.98 \pm 0.08}$ suggesting a common origin. This is supported by the OIR, X-ray and OIR to X-ray spectral indices being consistent with a single power law ($\alpha = -0.73$). Ostensibly the compact, synchrotron jet could therefore account for $\sim 100$ per cent of the X-ray flux at low luminosities in the hard state. At the same time, (i) an excess is seen over the power law decay of the X-ray flux at the point in which the jet would start to dominate, (ii) the X-ray spectrum slightly softens, which seems to be due to a high energy cut-off or break shifting to a lower energy, and (iii) the X-ray rms variability increases. This may be the strongest evidence to date of synchrotron emission from the compact, steady jet dominating the X-ray flux of an X-ray binary. For \1550, this is likely to occur within the luminosity range $\sim (2 \times 10^{-4} $ -- $ 2 \times 10^{-3})$ $L_{\rm Edd}$ on the hard state decline of this outburst. However, on the hard state rise of the outburst and initially on the hard state decline, the synchrotron jet can only provide a small fraction ($\sim$ a few per cent) of the X-ray flux. Both thermal Comptonization and the synchrotron jet can therefore produce the hard X-ray power law in accreting black holes. In addition, we report a phenomenonological change in the OIR spectral index of the compact jet from possibly a thermal distribution of particles to one typical of optically thin synchrotron emission, as the jet increases in energy over these $\sim 20$ days. Once the steady jet is fully formed and the infrared and X-ray fluxes are linearly correlated, the spectral index does not vary (maintaining $\alpha = -0.7$) while the luminosity decreases by a factor of ten. These quantitative results provide unique insights into the physics of the relativistic jet acceleration process.
\end{abstract}


\section{Introduction}

Numerous efforts have been made in recent years to identify the emission from jets in X-ray binary systems. These jets, produced close to accreting black holes and neutron stars, are known to (at least in some cases) carry a significant fraction of the accretion energy away from the binary, in the form of relativistic flows \citep*[e.g.][]{miraet92,gallet03,gallet05}. Steady, continuously replenished `compact' jets are seen in the hard X-ray state \citep[e.g.][]{fend06}. Like some jets produced by supermassive black holes in Active Galactic Nuclei (AGN), the emission is assumed to originate from synchrotron radiation produced by electrons or positrons (leptons) in the stratified jet.

From radio through infrared the observed radiation is a $\sim$ flat self-absorbed optically thick synchrotron spectrum with spectral index $\alpha \approx 0.0$ to +0.2 (where $F_{\nu} \propto \nu^{\alpha}$), composed of a superposition of synchrotron-emitting particle distributions \citep{blanko79}. The higher energy synchrotron emission originates in a small, dense region of the jet, close to the location where the jets are launched near the compact object \citep{blanko79,kais05}. Since $\alpha \simgt 0$, the bulk of the radiative power of the jet resides in this higher energy emission; at some frequency \citep[likely in the infrared; e.g.][]{corbfe02,russet06} there is a break in the jet spectrum from one which is $\sim$ flat ($\alpha \approx 0.0$ to +0.2) to optically thin (with $\alpha \approx -0.7$). In addition there is a cut-off in the jet spectrum at higher energies \citep*[likely in the X-ray regime; see e.g.][]{market01,maitet09}. Emission from the star and outer accretion disc often dominate the optical/infrared light, so the frequency of the aforementioned optically thick--thin jet break is hard to identify. Similarly, the inner accretion disc and hot inner flow/`corona' produces the majority of the X-rays \citep[for reviews see][]{charco06,gilf09}. However, in some cases emission from compact jets has been successfully isolated in the infrared, optical \citep[e.g.][]{corbfe02,buxtba04,miglet06} and possibly X-ray \citep*[e.g.][]{hyneet03} regimes. Moreover, it was shown \citep*{market05} that the base of the jet and the `corona' could be synonymous and the hard X-rays could arise from inverse Compton emission at the jet base. In addition, the optical emission from the jet is correlated with that of inflowing matter, providing information about how the two are coupled, or how the disc feeds the jet \citep*[e.g.][]{kanbet01,malzet04}.

An empirical unification of jet--disc coupling in black hole X-ray binaries (BHXBs) has been proposed. A BHXB usually traces out a hysteretical pattern in the X-ray hardness--intensity diagram \citep[HID; a similar hysteresis has been noted in accreting neutron stars and even white dwarfs;][]{maccco03,kordet08}, and the broadband behaviour is correlated with the evolution through the HID \citep*{donegi03,fendet04,doneet07,dunnet08,dunnet09,fendet09,cabaet09,bell09}.
Generally, the steady, compact jets exist in the hard state and are suppressed in the soft state \citep{gallet03,fendet04,fendet09}.

Since its discovery, the BHXB \1550 has performed outbursts or re-brightenings in 1998--99, 2000, 2001, 2002 and 2003 \citep{oroset02,bellet02,arefet04,dunnet09}. The compact object was found dynamically to be a black hole of mass $\sim 8$--12 $M_{\odot}$  \citep{oroset02}. The system is most famous for its arcmin-scale radio and X-ray jet `blobs' which were seen to decelerate several years after the jets were launched, due to jet--ISM interactions \citep{corbet02}. The outburst of \1550 in 2000 had very well-sampled optical, near-infrared (NIR) and X-ray monitoring throughout the entire outburst \citep*{jainet01,tomset01,reilet01,rodret03}. Here we analyse the light curves and spectral energy distributions (SEDs) and show that for this outburst it is possible to isolate the disc and jet components of the optical/NIR (OIR) emission. We use this to correlate changes in the broadband spectrum of the jet with evolution of the X-ray hysteresis. In Section 2 we describe the data collection. In Section 3 the multi-wavelength light curves and spectral evolution are analysed. The OIR jet emission is isolated and the broadband evolution of the jets are discussed. We constrain the contribution of the jet to the X-ray flux and plot the broadband SEDs. The results and implications are discussed in Section 4 and the conclusions are summarised in Section 5.

\begin{figure*}
\centering
\includegraphics[width=16cm,angle=270]{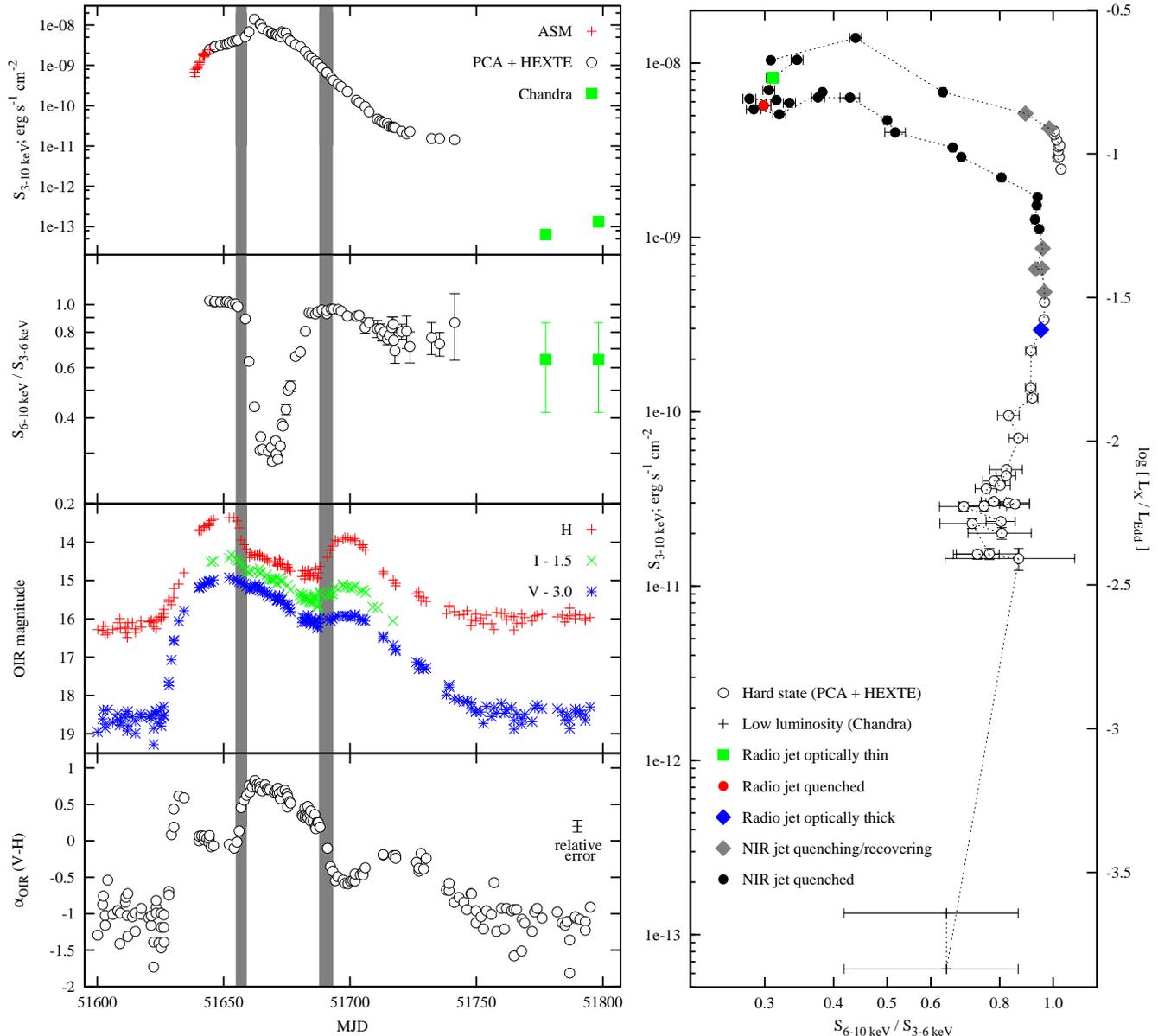}\\
\caption{Light curves (left) and X-ray HID (right) of the 2000 outburst of XTE J1550--564. The vertical, shaded (grey) regions in the light curves and the grey diamonds in the HID indicate when the OIR jet is quenching/recovering. From upper to lower panels on the left: X-ray 3--10 keV flux from the RXTE {\it PCA} + {\it HEXTE} spectral fitting prescription of \citeauthor{dunnet09} (\citeyear{dunnet09}), the {\it ASM} and Chandra \citep{tomset01} data (error bars for the {\it PCA} + {\it HEXTE} and Chandra data are smaller than the symbols); X-ray hardness ratio from {\it PCA} + {\it HEXTE} spectral fitting data and Chandra data (these are also used to construct the HID in the right panel); OIR apparent magnitudes from the YALO telescope \citep[][errors are $< 0.05$ mag]{jainet01}; and the OIR intrinsic spectral index derived from $V$ optical and $H$ NIR de-reddened flux densities (using $A_{\rm V} = 5.0$; see text).}
\end{figure*}

\section{Data Collection}

Data from the Rossi X-ray Timing Explorer (RXTE) Proportional Counter Array ({\it PCA}), High Energy X-ray Timing Experiment ({\it HEXTE}) and All-Sky Monitor ({\it ASM}) were used. We adopt the {\it PCA} and {\it HEXTE} X-ray spectral fitting methodology of \cite{dunnet09}. For the full details we refer the reader to \cite{dunnet09}. This procedure resulted in a total number of fitted, usable observations in the archive of 62 within the time period of interest in this study.

The {\it ASM} fluxes are used only on the (hard state) rise of the outburst, before the pointed {\it PCA} observations were triggered. The 3--10 keV {\it ASM} fluxes are estimated from the {\it ASM} 1.5--12 keV count rates using the \emph{NASA} tool \emph{Web-PIMMS} assuming a power law photon index of $\Gamma = 1.6$ (where $\Gamma = 1 - \alpha$; 1.6 is typical for BHXBs in the hard state; this index is also that measured by the {\it PCA} during the hard state rise of this outburst just after the {\it ASM} data that are used here). In addition, two Chandra detections after the main outburst are taken from \cite{tomset01}. The 3--10 keV Chandra fluxes are estimated from the quoted 1.5--12 keV fluxes and quoted $\Gamma = 2.3$ \citep[measured from the Chandra data, by][]{tomset01}. To estimate luminosities as fractions of the Eddington luminosity $L_{\rm Edd}$, we adopt a distance of 5.3 kpc to the source and a black hole mass of 10.6 $M_{\odot}$ \citep{oroset02}. We approximate the bolometric luminosity to 5.0 times the 2--10 keV X-ray luminosity, which is appropriate for BHXBs in the hard state \citep{miglfe06}.

We take apparent V (0.55 $\mu$m) and I (0.81 $\mu$m) optical and H (1.7 $\mu$m) infrared waveband magnitudes from \citeauthor{jainet01} (\citeyear{jainet01}; obtained using the YALO telescope) and radio flux densities from \citeauthor{corbet01} (\citeyear{corbet01}; from observations made with the Australia Telescope Compact Array). The OIR data were also used in several follow-up works \citep{corbet01,oroset02,russet06,yuanet07,russet07,russet08}. To derive intrinsic, de-reddened OIR flux densities, we use $A_{\rm V} = 5.0$ (\citealt{tomset01,tomset03,kaaret03}; see also discussion in \citealt{russet07}) for the optical extinction and adopt the extinction law of \cite*{cardet89}.

\section{Data Analysis}

\subsection{Outburst evolution}

In Fig. 1 we present the X-ray and OIR light curves of the 2000 outburst of \1550 \citep[left panel; see also][]{tomset01,jainet01,corbet01,reilet01,rodret03}. The evolution of the X-ray hardness and OIR colour (spectral index $\alpha$, in this case between V and H-bands) are also plotted. The source made a transition from a hard state to a soft/intermediate state around MJD $\sim$ 51655--65 \citep[in the context of jet activity the source was in a soft state; see discussion in][]{russet07} and back, at lower luminosity, to a hard state around MJD $\sim$ 51675--85 \citep[e.g.][]{rodret03}, obeying the well known BHXB hysteresis in the HID \citep[e.g.][]{fendet09}. There is a quenching and recovering of the NIR flux (and the optical to a lesser extent) contemporaneously with the X-ray softening and hardening, respectively \citep[see also][]{jainet01}. The periods in which the NIR flux is fading or recovering are shaded grey in the light curves (MJD $\sim$ 51655--59 and $\sim$ 51688--93). Synchrotron emission from the jet accounts for the NIR flux during the hard state, this jet quenching during the soft state \citep[e.g.][]{jainet01,corbet01}. \cite{russet07} showed that the NIR emission from the jet is brighter during the hard state decline than during the hard state rise, at a given X-ray luminosity (NIR--X-ray hysteresis), which may be due to a stronger jet on the decline, or changes in the jet radiative efficiency or disc viscosity.

In the right panel of Fig. 1 we present the X-ray HID; information regarding the radio and NIR jet are indicated. HIDs of this outburst are also presented in \cite{reilet01,rodret03,gierne06,dunnet09}.
We see that the NIR jet starts to drop as soon as the source begins to soften and leaves the hard state \citep[the same behaviour as that seen in GX 339--4;][]{homaet05}. The NIR jet does not reappear until the source is back in the hard state and declining in luminosity \citep[the same as was observed for 4U 1543--47 and GX 339--4;][]{buxtba04,coriet09}. A radio jet was present when the X-ray spectrum was much softer (on MJD 51665; shown by the green point in the X-ray HID) but then was quenched when the source decreased in luminosity in the soft state (MJD 51670; red point). This radio detection had an optically thin spectrum and probably originated in discrete ejecta released from the core at a previous epoch \citep{corbet01,rodret03}. The radio jet, with an optically thick spectrum (consistent with originating from the compact, self-absorbed jet) was then detected in the hard state once the NIR jet had recovered (MJD 51697; blue point). No radio data were taken prior to this during the soft-to-hard transition, so it is uncertain when the radio jet reappeared. As far as we are aware, no other radio data of this outburst exist in the literature.

\begin{figure}
\centering
\includegraphics[width=\columnwidth,angle=0]{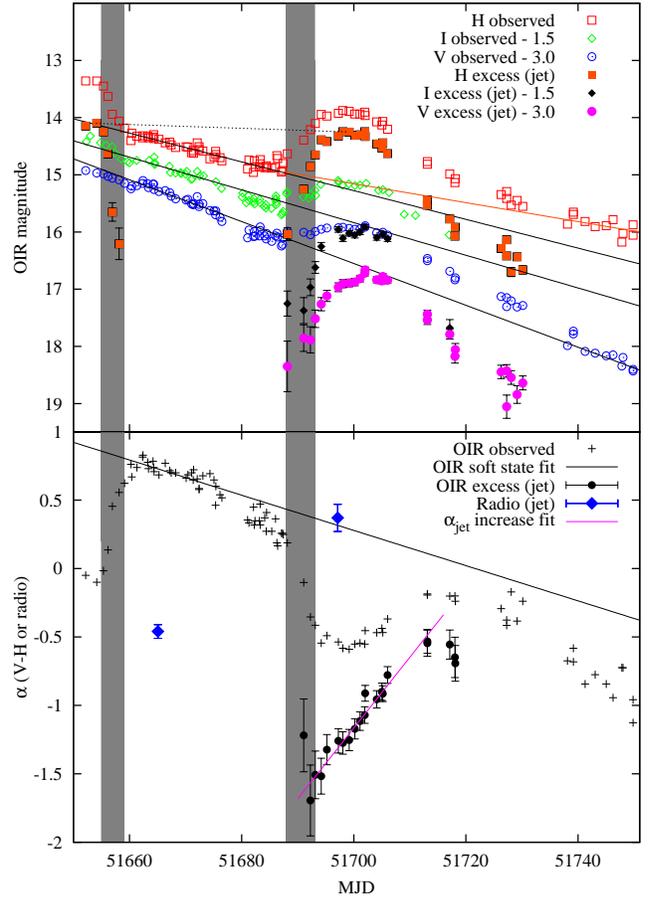}
\caption{Light curve and OIR spectral index of the different emitting components in XTE J1550--564 during the 2000 outburst. \emph{Upper panel:} The OIR magnitudes are apparent (not de-reddened) and the linear fits (black solid lines) to the fading in the soft state are 0.025, 0.029 and 0.037 mag day$^{-1}$ for H, I and V-bands, respectively. The V-band data return to this fit after the excess from the jet has become negligible again (MJD $\sim$ 51740) whereas the H-band has already reached the quiescent level of $H \sim 16.0$ and so does not return to this fit because it is fainter than the quiescent level (although we do assess the effect of a different H-band slope as shown by the orange line). The almost horizontal dotted line helps to emphasise the very similar peak fluxes of the jet in the hard states before and after the soft state (the two flux peaks of the black squares). \emph{Lower panel:} The spectral indices $\alpha_{\rm OIR}$ are intrinsic (derived from de-reddened flux densities). Radio data are from \citeauthor{corbet01} (\citeyear{corbet01}). The fit to the smooth change in spectral index of the excess from the jet (magenta line) is $\alpha$ increasing by 0.052 day$^{-1}$.}
\end{figure}

\subsection{Isolating the jet emission}

The OIR light curves during the soft/intermediate state (while the viscous disc and/or the irradiated disc dominated) resemble power law decays. We measure a V, I and H-band fade of 0.037, 0.029 and 0.025 mag per day respectively, during this time (MJD 51660--51680) which are very similar to the values derived by \cite{jainet01}. By extrapolating these power laws we can estimate the V and H magnitudes of the excess (jet) component, and therefore the spectral index of the jet, in the return to the hard state. We approximate the jet flux in each waveband to be the excess flux above the soft state power law decay in that waveband (Fig. 2). We remark that no abrupt changes in the OIR flux or spectrum are seen over the state transitions (such changes may be expected from an irradiated disc due to changes in the X-ray spectrum) but instead changes are seen gradually over timescales of days, after the X-ray transition is complete. \cite{jainet01} fit the light curves with power law decays plus Gaussian profiles approximating the jet rise and decline; here we assume only the underlying power law decay and measure the excess of the jet recovery above this decay. The power law decay of the disc component likely continued with the same slope; once the jet faded the V-band light curve rejoined the power law around MJD $\sim$ 51740 with the same slope and normalisation (Fig. 2, upper panel). The H-band light curve is nearing its quiescent level and so cannot fade further to rejoin the power law. The companion star has a redder spectrum than the disc, and likely dominated the H-band while the V-band was still dominated by the disc. We do however show an alternative slope for the H-band disc decay and evaluate the impact of this different slope (see below). The extrapolation of the soft state power law is also done for the hard state just before the soft state, but only in H-band because there is a pronounced drop in the H-band flux during transition to the soft state, and no such drop in the I and V-bands.

\begin{figure}
\centering
\includegraphics[height=\columnwidth,angle=270]{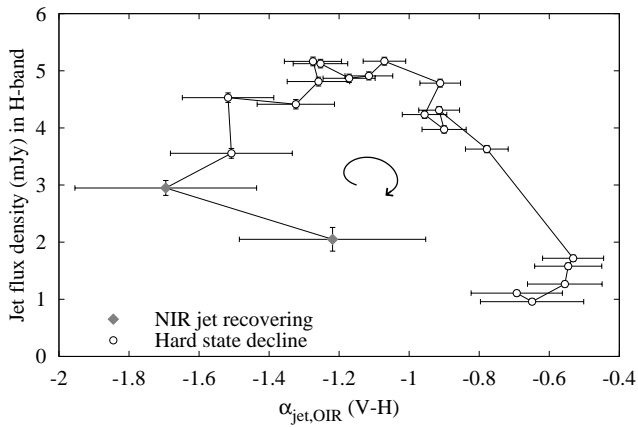}
\caption{The evolution of the OIR spectral index $\alpha$ of the jet as it rises and fades in the hard state decline of the outburst. The symbols are the same as those in the HID in the right panel of Fig. 1. The errors shown are relative; there may be systematic errors in $\alpha$ from the uncertainty in the optical extinction.}
\end{figure}

The first thing to note is that the peak of the jet flux in H-band is very similar on the hard state decline compared to on the hard state rise of the outburst (see the orange squares and the black dotted line in the upper panel of Fig. 2). The NIR jet peaked at almost identical flux levels, suggesting that on the outburst decline, the jet `remembers' how bright it was on the hard state rise. On the hard state rise, the X-ray and IR jet were both brightening together, whereas on the hard state decline it took some time for the IR jet to reach peak flux (the same peak flux level as before the soft state), but at the same time the X-ray flux was already fading. This caused the NIR--X-ray hysteresis behaviour discussed in \cite{russet07}, but the reason for the IR jet to peak at the same flux level on both rise and decay remains unknown. We note that for GX 339--4 (an other well studied BHXB) the NIR jet does not rise to the same flux level on the hard state decline, and NIR jet--X-ray hysteresis is not observed \citep{coriet09}. A radio--X-ray hysteresis is seen for GX 339--4 however (Corbel et al. in preparation).

We find that as the NIR jet of \1550 brightened across the transition from the soft back into the hard state, the spectral index of the NIR jet was steep: $\alpha_{\rm jet} \approx -1.5$ and then increased smoothly during the next $\sim 20$ days in the hard state (MJD $\sim$ 51692--712; Fig. 2, lower panel). We measure $\alpha$ to increase at a rate of 0.052 per day (the fit is shown by the magenta line in the lower panel of Fig. 2). The index then remained at a value of $\alpha_{\rm jet} \approx -0.6$ to $-0.7$ as the source faded, typical of optically thin synchrotron emission. These values of $\alpha_{\rm jet}$ may suffer from systematic uncertainties propagating from the error in the interstellar extinction, but the smooth change in $\alpha_{\rm jet}$ with time appears to be real. The optically-thick to optically-thin break in the jet spectrum for this outburst must reside at frequencies lower than H-band, in the mid-infrared.

The spectral index of optically thin synchrotron emission is dependent on the lepton energy distribution $p$ in the flow; $\alpha_{\rm jet} = (1-p)/2$. If the emission is optically thin since the start of the rise of the NIR jet, the change in $\alpha_{\rm jet}$ represents a change in $p$, from $p \sim 4.0$ to $p \sim 2.2$. $p \sim 4.0$ is very steep and may well be an exponential cut-off, so the jet plasma may be gradually switching from a thermal, possibly Maxwellian distribution of particles to a non-thermal distribution becoming more dominant as $p$ increases. We therefore may be witnessing an intrinsic change in the distribution of particle (lepton) energies in the flow of the steady, compact jet over these 20 days. The implication is that the physics of the acceleration process at the jet base is changing with time; the jet is gradually becoming more energetic; a gradual re-ignition of the jet. In Fig. 3 it is demonstrated that as $\alpha$ increases, the NIR jet flux first rises then falls in the decline of the outburst. The value of $\alpha_{\rm jet}$ only stabilizes to $\alpha_{\rm jet} \approx -0.6$ once the jet flux is already fading. We cannot measure $\alpha_{\rm jet}$ from the OIR in the hard state rise of the outburst because the V-band magnitude did not fade considerably during the hard-to-soft transition (unlike the H-band, which did), implying a low jet contribution to the V-band in the hard state rise.

At the time of the gradual increase in $\alpha_{\rm jet}$ in the OIR, the radio (1--9 GHz) jet was already optically thick \citep[Fig. 2;][]{corbet01}. This is also close in time to the peak NIR jet flux. The radio and H-band flux densities of the jet at this time were $\sim 1$ mJy and $\sim 4$ mJy, respectively; the radio-to-NIR spectral index was $\alpha_{\rm jet} = +0.15$; also fairly typical for optically thick jets \citep[e.g.][]{fend01,russet06}. The OIR SEDs of the separated disc and jet components are presented in Fig. 4. The SEDs of the disc (blue dotted lines at the top of the figure) are from during the soft state when no jet was detected, and are generally blue, with $\alpha_{\rm disc} \sim +0.7$ between V and H-bands. This is typical of irradiated discs of BHXBs in outburst \citep[e.g.][]{hyne05} but could also be the viscously heated disc \citep[although see][Russell, Maitra et al. in preparation]{russet08}. The brightest flux density of the disc appears to reside in the I-band, which could occur if the peak of the blackbody spectrum is between the V and H-bands (in the SEDs of \citealt{hyne05} this peak is generally at higher frequencies). Conversely, the effect may simply be due to inaccuracies in the extinction. The SEDs of the jet component are clearly redder as the jet rises into the hard state (shown by the black dashed lines in Fig. 4) and bluer as the jet fades (red solid lines). The arrows in the right hand side of the plot indicate the evolution of the respective SEDs. When the NIR jet is close to its brightest, it is detected in radio; the radio to NIR spectral index of the jet is indicated on the left hand side of the figure.

\begin{figure}
\centering
\includegraphics[width=\columnwidth,angle=0]{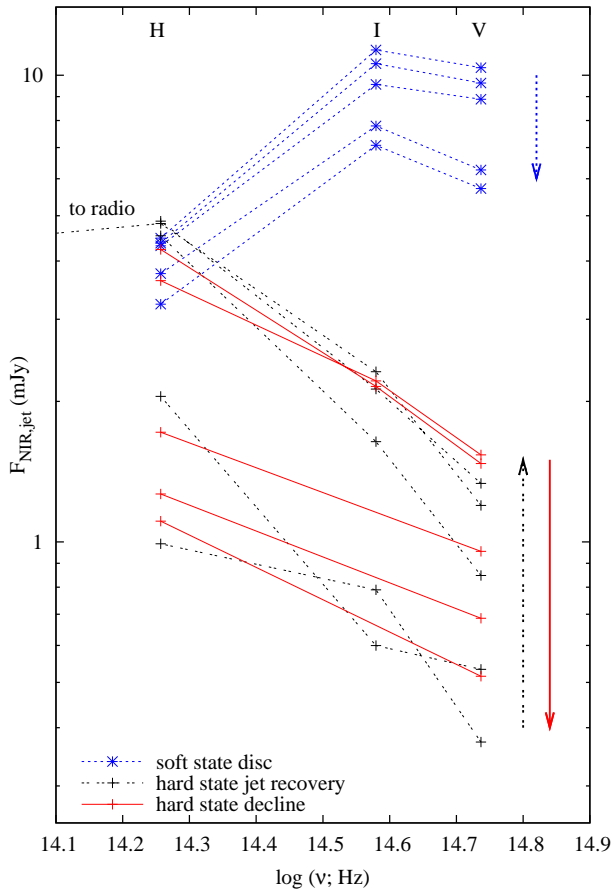}
\caption{OIR SEDs of the isolated disc in the soft state (blue data; dotted lines at the top) and jet (black dashed lines for the jet recovery into the hard state and red solid lines for the hard state decline).}
\end{figure}

\begin{figure}
\centering
\includegraphics[height=\columnwidth,angle=270]{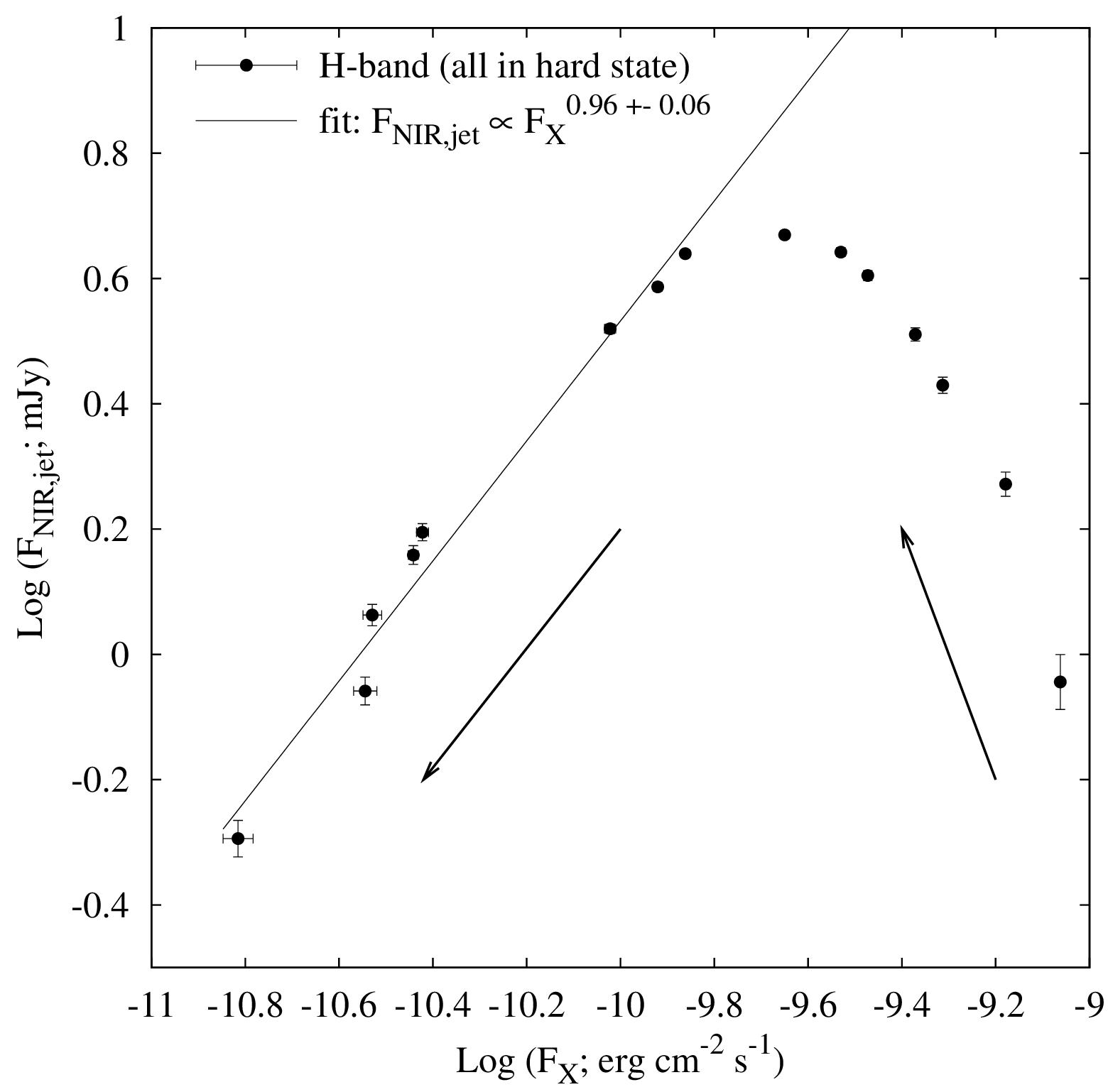}
\caption{NIR--X-ray correlation on the decline of the XTE J1550--564 outburst, when the H-band emission from the jet only is used (the trend from the disc is subtracted). The NIR emission from the optically thin jet and the X-ray emission are directly proportional once both are fading in the hard state (after MJD 51702). This suggests that both the OIR and X-rays originate in the optically thin part of the jet synchrotron spectrum.}
\end{figure}

\subsection{The NIR jet--X-ray correlation}

From modelling compact jets it was shown that the total jet luminosity scales linearly with the mass accretion rate, $L_{\rm jet} \propto$ \.m and the radio jet luminosity varies as $L_{\rm Radio,jet} \propto L_{\rm jet}^{1.4} \propto$ \.m$^{1.4}$ \citep{falcbi95,falcbi96,heinsu03}. Similarly the X-ray flux is expected to correlate with the mass accretion rate as $L_{\rm X}\propto$ \.m$^{\sim 2}$ for black holes in a hard state \citep[the accretion process is radiatively inefficient; e.g.][]{narayi95,kordet06b}. Coupling these two relations leads to $L_{\rm Radio,jet} \propto L_{\rm X}^{0.7}$, and $L_{\rm OIR,jet} \propto L_{\rm Radio,jet} \propto L_{\rm X}^{0.7}$ if the radio to OIR jet spectrum is optically thick. These relations hold if the X-ray emission originates in the hot (inflowing) accretion flow (hence jet--disc coupling) or the jet synchrotron emission. From observations of compilations of sources, these correlations have been found to describe the data fairly well for BHXBs in the hard state \citep{falcbi96,fend01,corbet00,corbet03,market03,gallet03,gallet06,russet06}. The radio--X-ray correlation in particular is extended (with a mass term) to include AGN, providing strong evidence for a similar disc--jet coupling mechanism for all accreting black holes \citep*[the `fundamental plane of black hole activity'; e.g.][]{merlet03,falcet04,kordet06a,gultet09}.

\begin{figure}
\centering
\includegraphics[width=\columnwidth,angle=0]{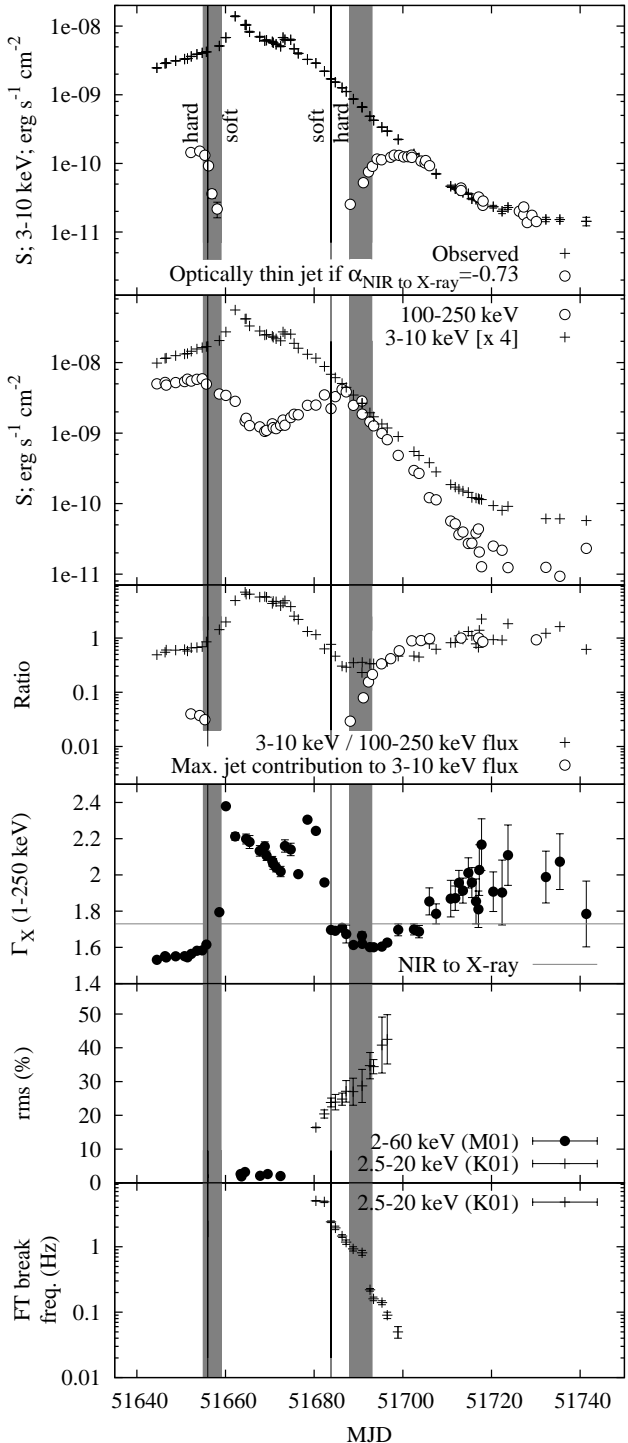}
\caption{X-ray light curves, `hard' hardness ratio and possible contribution from the optically thin jet extrapolated from the OIR (upper three panels) and light curves of the measured X-ray photon index, the rms variability and the break frequency in the power spectrum (lower three panels; the latter two are taken from M01 = \citealt{millet01} and K01 = \citealt{kaleet01}).}
\end{figure}

To investigate whether the OIR jet indeed correlates with the X-ray flux we plot one against the other in Fig. 5. The intrinsic flux density of the jet in H-band (achieved by subtracting the trend of the disc from the light curve; see Fig. 2) is plotted against the 3--10 keV X-ray flux. Once the jet has risen during the transition to the hard state and starts to fade, the NIR jet and X-ray fluxes are tightly linearly correlated, with $L_{\rm NIR,jet} \propto L_{\rm X}^{0.96 \pm 0.06}$. We note that if the different H-band soft state decay (orange line in Fig. 2) is invoked, the correlation changes to $L_{\rm NIR,jet} \propto L_{\rm X}^{1.01 \pm 0.05}$ if all but the lowest data point are included, and $L_{\rm NIR,jet} \propto L_{\rm X}^{1.13 \pm 0.11}$ if the lowest data point is included. This lowest data point is when the H-band flux from the jet is only $\sim 0.1$ mag above the apparent extrapolated disc flux at that epoch, which is comparable to the H-band scatter, so we exclude it due to its very large error. We therefore measure the correlation, taking into account the uncertainty in the H-band disc decay, as $L_{\rm NIR,jet} \propto L_{\rm X}^{0.98 \pm 0.08}$ (this range encompasses the extremes of both fit errors).

This is indeed steeper than expected from the above relations; the two are directly proportional. Previously, the NIR--X-ray correlation in this source was measured to be $L_{\rm NIR} \propto L_{\rm X}^{0.63 \pm 0.02}$ from the same data \citep{russet07} but in this instance the jet emission had not been isolated from the disc. The correlation slope of $0.63 \pm 0.02$ is true for the total NIR versus X-ray emission on the hard state decline; the $0.98 \pm 0.08$ slope is true for the NIR jet emission versus X-ray emission, after the disc flux has been subtracted from the NIR flux. This demonstrates the importance of isolating the OIR emission contributions if OIR--X-ray correlation slopes are to be interpreted. The tight linear correlation exists over one order of magnitude in flux, and implies the two fluxes may share the same origin. In fact, the OIR spectrum is optically thin, so we would expect $L_{\rm OIR,jet} \propto L_{\rm X}$; a linear correlation if the X-ray flux also originates in the optically thin synchrotron jet. This possibility is investigated in Section 3.4.

\subsection{Synchrotron jet contribution to the X-ray flux}

Since the NIR jet flux and X-ray flux are directly proportional on the hard state decline, it is worth assessing whether the flux from this synchrotron emission from the jet could account for the X-ray flux itself \citep[see e.g.][]{market01,market03,nowaet05}. We find that the spectral index between NIR jet and observed X-ray flux is $\alpha = -0.73$ when the NIR and X-ray fluxes are linearly correlated (MJD 51702 -- 51730). Independently, we measure the OIR (H to V) spectral index of the jet on these dates to be $\alpha = -0.76 \pm 0.23$, and the X-ray power law spectral index from spectral fitting to be initially $\alpha = -0.70 \pm 0.03$ (photon index $\Gamma_{\rm X} = 1.70 \pm 0.03$ then softening to $\alpha = -1.1 \pm 0.2$ ($\Gamma_{\rm X} = 2.1 \pm 0.2$). Since these three independent measurements of $\alpha$ (in OIR, in X-ray and the OIR to X-ray spectral index) are consistent with being the same value (at least initially), the whole spectrum appears to likely be one single power law declining in luminosity. If this is the case, the synchrotron jet would account for $\sim 100$ per cent of the X-ray flux.

The 3--10 keV X-ray light curve is again plotted in Fig. 6 (upper panel), with the expected contribution from the optically thin jet in this band, if $\alpha = -0.73$ is extrapolated from NIR to X-ray, overplotted. The two light curves overlap exceptionally well after MJD $\sim$ 51700. Before this time, a power law decay is evident in the X-ray light curve \citep[in the soft state and continuing into the hard state; see also Fig. 1 of][]{tomset01}, but after this time an excess is present above the power law, which is consistent with the optically thin synchrotron emission from the jet. Thirty six days after the last X-ray detection in our light curve, \cite*{corbet06} measured an unabsorbed X-ray flux of $3 \times 10^{-13}$ erg s$^{-1}$ cm$^{-2}$ (0.5--10 keV) with Chandra on MJD 51777, almost two orders of magnitude fainter than the flux 36 days before (see also Fig. 1). This implies the excess above the power law decay from MJD $\sim 51700$ onwards is not the BHXB simply levelling off to a constant, quiescent level; instead the source later faded further by almost two orders of magnitude. Background counts within the PCA field were negligible because only PCA data with more than 1000 background-subtracted counts were fitted \citep{dunnet09}. The spectral index of $\alpha = -0.73$ is typical for optically thin synchrotron emission.

In the second panel of Fig. 6 we plot the high energy X-ray flux (100--250 keV) and the soft X-ray flux (3--10 keV) light curves, and in the third panel the flux ratio of these two X-ray bands, along with the maximum jet contribution to the X-ray flux (derived above). We see from this panel that on the hard state rise of the outburst, the synchrotron jet flux as extrapolated from NIR could not have contributed more than a few per cent of the X-ray flux. Similarly on the hard state decline of the outburst, the jet contributed $\simlt 3$ per cent of the X-ray flux initially, but by MJD 51702 it could provide $\sim 100$ per cent of the X-ray flux. The high energy X-ray flux appears to fade faster than the soft X-rays at this point (in fact it approximately continues the power law decay extrapolated from the soft state), softening the X-ray spectrum. This could either represent a steepening in the slope of the power law, or a decrease in the high energy cut-off or break in the spectrum to one at a lower energy. The slope of the NIR to soft X-ray power law remained the same during this time, but the high energy X-ray flux decreased. This implies the high energy cut-off or break in the power law shifted to a lower energy when the jet made its maximum contribution to the X-ray flux. This is supported by the spectral fits of \cite{rodret03}; these authors detect a cut-off at $\sim 60$ keV when the source re-entered the hard state, which decreased to $\sim 30$ keV and then is no longer necessary in their fits after MJD 51690, which is when the jet likely began to contribute significantly. BHXBs in the hard state are generally known to have a cut-off in the X-ray power law at around 100 keV \citep[e.g.][]{gilf09}. Here, we see a possible shift in the cut-off to lower energies at the same time as the synchrotron jet possibly begins to dominate the X-rays.

In the lower three panels of Fig. 6 we plot a number of additional observables: the measured X-ray power law photon index $\Gamma_{\rm X}$ (fitted from combined {\it PCA} and {\it HEXTE} data), the rms variability measured by \cite{millet01} and \cite{kaleet01}, and the frequency of the break in the Fourier Transform (FT) power spectrum \citep{kaleet01}. As the jet begins to possibly dominate the X-ray flux, a clear increase in $\Gamma_{\rm X}$ from 1.6 to 1.7 and then finally to $\geq 1.8$ is observed. An increase in the rms variability and a decrease of the FT break frequency are also noted, however these variables are changing before MJD $\sim 51690$, when the jet contributed less than a few per cent of the X-ray flux. We would expect the synchrotron emission from the jet to display rapid, high amplitude variability, as is seen in the OIR \citep[e.g.][]{hyneet03}. The jet could not have caused the increase in rms (and decrease in the FT break frequency) seen before $\sim 51690$, but could contribute to the increased rms between MJD $\sim 51690$ and $\sim 51700$.

\begin{figure}
\centering
\includegraphics[height=\columnwidth,angle=270]{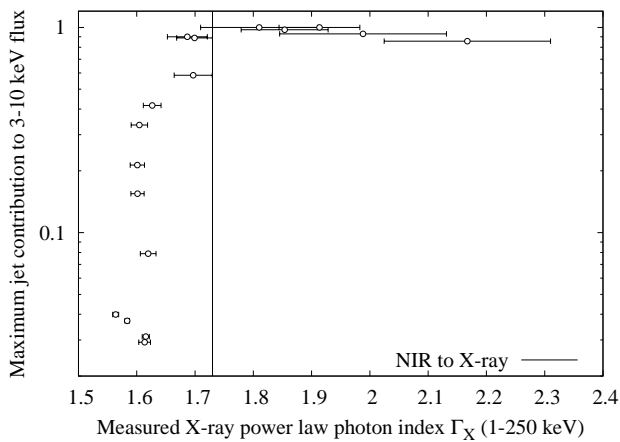}
\caption{The maximum contribution the synchrotron jet can make to the X-ray 3-10 keV flux, plotted against the measured X-ray power law photon index. Once the maximum jet contribution rises above 60 per cent of the X-ray flux, the power law changes from $\Gamma = 1.6$ to $\Gamma = 1.7$, very similar to the measured NIR (jet) to X-ray power law (indicated by the vertical grey line). When the jet could contribute $\approx 100$ per cent of the X-ray flux, the measured X-ray power law has a photon index of $\Gamma = 1.8$--2.2.}
\end{figure}

$\Gamma_{\rm X}$ is plotted against the maximum jet contribution to the X-ray flux in Fig. 7. We find that $\Gamma_{\rm X} \approx 1.6$ until the point at which the jet contributes up to 60 per cent of the flux. $\Gamma_{\rm X}$ then steepens to 1.7, very close to the measured NIR to X-ray spectral index of the jet, 1.73 (indicated by the vertical line in Fig. 7). This supports the suggestion that the synchrotron jet at this point dominates the X-ray flux. $\Gamma_{\rm X}$ is then seen to steepen further, to values of $\approx 1.8$--2.2. This could occur if the high energy cut-off in the synchrotron spectrum of the jet moves to energies around 3--10 keV as the jet becomes less energetic \citep[this is consistent with the decrease in energy of the cut-off just before this epoch as fit by][]{rodret03}. The high energy (100--250 keV) flux continues a power law fade in the light curve, unlike the soft (3--10 keV) flux (Fig. 6, second panel). This is consistent with a cut-off in the jet spectrum residing at $< 100$ keV, the jet making a negligible contribution to the flux above 100 keV. Alternatively, the high energy cut-off could reside at much lower energies than 3--10 keV, and the jet could contribute a negligible amount of the X-ray flux. If this were the case, the linear NIR--X-ray correlation, the X-ray excess over the power law decay and the apparently single power law slope from NIR to X-ray would all have to be coincidental. We cannot rule this out, but the data certainly favour the synchrotron emission from the jet producing the majority of the X-ray flux after MJD $\sim 51700$. In addition, we show in Section 4.1 that the major arguments usually used to reject the synchrotron jet as the origin of the X-ray flux in the hard state of BHXBs are not valid for this case of \1550.

\begin{figure}
\centering
\includegraphics[height=\columnwidth,angle=270]{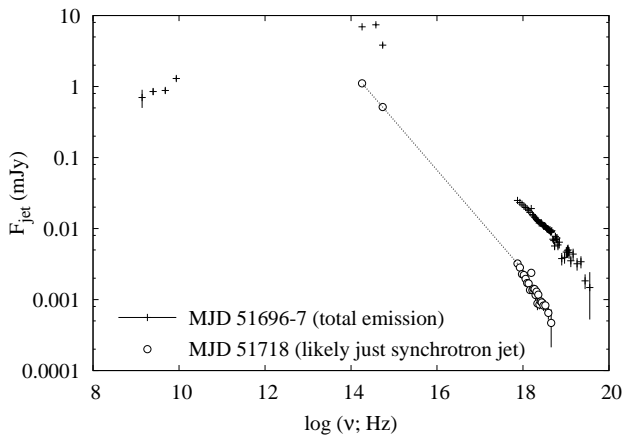}
\caption{Broadband SEDs of \1550. On MJD 51696--7 we show the entire, intrinsic radio through X-ray SED (the X-ray data were unabsorbed assuming an additive continuum model; see text) which probably includes emission from the disc, the jet and thermal Comptonization. On MJD 51718 the jet emission has been isolated; we estimate $\sim$ zero contamination from other components. The jet likely contributes $\sim 40$ and $\sim 100$ per cent of the X-ray flux on MJD 51696--7 and MJD 51718, respectively (Fig. 6).}
\end{figure}

\subsection{Broadband SEDs}

In Fig. 8 we plot two unabsorbed (intrinsic) broadband SEDs from radio to X-ray of \1550 during the hard state. The X-ray data were corrected for photoelectric absorption assuming a phenomenological multicoloured accretion disk plus power law continuum model. The first SED is from the hard state decline soon after the transition (MJD 51696--7), when the radio SED was obtained. At this time, the jet likely contributed $\sim 40$ per cent of the X-ray flux (Fig. 6). For this reason we cannot isolate the X-ray jet spectrum in this SED; instead we plot the total SED as observed, which includes light from the disc in OIR and probably thermal Comptonization in X-ray, as well as the jet in these regimes and radio. The second SED is from MJD 51718, at a time in which the X-ray spectrum is likely dominated by the jet, at lower luminosity. Here we show the isolated jet spectrum (the disc has been subtracted from the OIR). Unfortunately no radio data were acquired at this time so we cannot fit the SED with a jet synchrotron model. Note the single power law from OIR to X-ray implied by the SED, and the slight softening of the X-ray flux compared to the SED at higher luminosity. The break between optically thick and optically thin emission in the jet spectrum lies at a frequency lower than $10^{14}$ Hz in both SEDs. The spectral indices of the optically thick and optically thin regimes are $\alpha \sim +0.15$ (derived from the first SED but after the disc emission has been subtracted) and $\alpha \sim -0.73$, respectively. The high energy cut-off in the jet spectrum may lie at $\sim 10^{18}$ Hz (around 3--10 keV; see Section 3.4).

\section{Discussion}

\subsection{Two separate X-ray power laws in the hard state}

The data between MJD 51702 and 51730 are consistent with being a single power law from the optically thin jet, with this component producing $\sim 100$ per cent of the 3--10 keV X-ray flux and a large fraction of the NIR flux. This result has major implications for the origin of the X-ray emission in the hard state of X-ray binaries. \cite{market01} originally suggested that optically thin synchrotron emission from the jet dominates the X-ray flux of the BHXB XTE J1118+480; in more recent, advanced models the synchrotron jet component produces typically $\sim 10$ per cent of the X-ray flux in the hard state \citep[e.g.][]{market05,miglet07}. For \1550 during the outburst in question, if the optically thin synchrotron emission does indeed dominate the X-rays, it does not do so until the source has faded to $S < 10^{-10}$ erg s$^{-1}$ cm$^{-2}$ (3--10 keV; Fig. 6), which is after the source has already faded by $> 1$ order of magnitude in the hard state decline. These results support the suggestion by \cite*{fendet03} that the energetics of BHXBs become jet dominated at low luminosities in the hard state. However, it does not appear that the synchrotron jet dominates the X-ray flux in the hard state rise of the outburst; this is not surprising since it was found that the NIR jet was fainter (at a given X-ray luminosity) on the hard state rise compared to the decline \citep{russet07}. This implies that two separate components can produce the X-ray power law in the hard state, the jet at low luminosities (at least for the decline of the 2000 outburst of \1550). The synchrotron-emitting jet then likely dominates the X-ray flux in the range $10^{-10} < S < 10^{-11}$ erg s$^{-1}$ cm$^{-2}$, or $\sim (2 \times 10^{-4} $ -- $ 2 \times 10^{-3})$ $L_{\rm Edd}$.

Traditionally, it is supposed that Compton upscattering of soft photons on hot electrons in a `corona' surrounding the compact object produces the hard X-ray power law in BHXBs (e.g. \citealt{sunyti80}; see \citealt{gilf09} for a review). This has been supported by the ability of Comptonization models to successfully reproduce observed X-ray SEDs in detail. It has also been shown \citep{market05} that inverse Compton emission at the base of the jet (partly fed by synchrotron photons also from the jet; this is sometimes referred to as synchrotron self-Compton) can also account for the hard X-ray spectrum of BHXBs. A number of arguments were put forward illustrating that the hard X-ray power law of BHXBs is unlikely to originate in the synchrotron jet itself.

It is claimed that the cut-off in the X-ray spectrum resides at $\sim 100$ keV for all BHXBs in the hard state \citep[within a factor of $\sim 2$; e.g.][]{grovet98}. \citep{zdziet03} argue that fine tuning of jet model parameters would be necessary to reproduce this ubiquity. For \1550 we do see an apparent evolution of the cut-off energy during times when the jet contribution to X-ray is changing. Fine tuning of jet parameters is no longer required for this source at this time. In addition, it was found \citep*{miyaet08,joinet08,mottet09} that the high energy cut-off varies between 40 keV and $> 200$ keV with changing luminosity in the hard state of GX 339--4 and between $\sim 100$ keV and $> 200$ keV in the hard state of GRO J1655--40.

\cite{macc05} estimated an expected increase of a factor 1000 in luminosity in transition to the soft state if the hard state emission was radiatively inefficient, based on 1 per cent of the accretion energy being channeled into the jet, which radiates 10 per cent of its energy. It has now been shown that jets may extract $\sim 50$ per cent of the accretion energy \citep[e.g.][]{gallet05}, which corresponds to an expected luminosity change of a factor 20 over the transition, much less than the factor 1000 estimated previously. The jet of \1550 seems to dominate the X-ray flux after one order of magnitude of fading in the hard state, which is approximately consistent with the predicted factor of 20. However, during this fading the mass accretion rate will probably have also decreased, so the jet may require a high radiative efficiency to be able to radiate so brightly in X-ray at $10^{-3}$ $L_{\rm Edd}$ (Tom Maccarone, private communication).

It was demonstrated \citep{hein04} that the X-ray spectral index must be steeper (a higher photon index $\Gamma_{\rm X}$) than is observed if the synchrotron jet produces the X-ray luminosity and this is responsible for the radio--X-ray--mass correlation in BHXBs and AGN (the `fundamental plane'; see Section 3.3). Similarly, it was proposed \citep{macc05} that jets cannot dominate the X-ray flux of BHXBs or neutron star X-ray binaries because their X-ray spectra in the hard state are very similar, implying a common emission origin, whereas the radio jets of BHXBs are much brighter than those of neutron star systems. While here we are not claiming the X-ray originates in the jet for all BHXBs in the hard state, we do find that the X-ray spectrum from the jet is very similar to that of the other hard state power law, and we do see a steepening of the X-ray spectral index when the jet may dominate \citep[in line with the theoretical prediction of][]{hein04}. In \cite{yuanet07}, it was argued that the jet contribution to the X-ray flux was low because an increase in X-ray flux was not observed in the hard state decline whereas an OIR increase from the jet was observed. Here we have shown that an increase in X-ray flux is not necessary; an excess above the X-ray power law decay is required only, and observed \citep[Fig. 6; see also][]{tomset01}.

To summarise, the arguments put forward to reject the synchrotron jet as a viable origin for the X-ray emission in BHXBs cannot be applied to \1550 on the hard state decline, below $2 \times 10^{-3}$ $L_{\rm Edd}$. In fact, we confirm that at the highest luminosities in the hard state (on both outburst rise and decline) the synchrotron jet cannot dominate the X-ray flux.

\begin{figure}
\centering
\includegraphics[width=\columnwidth,angle=0]{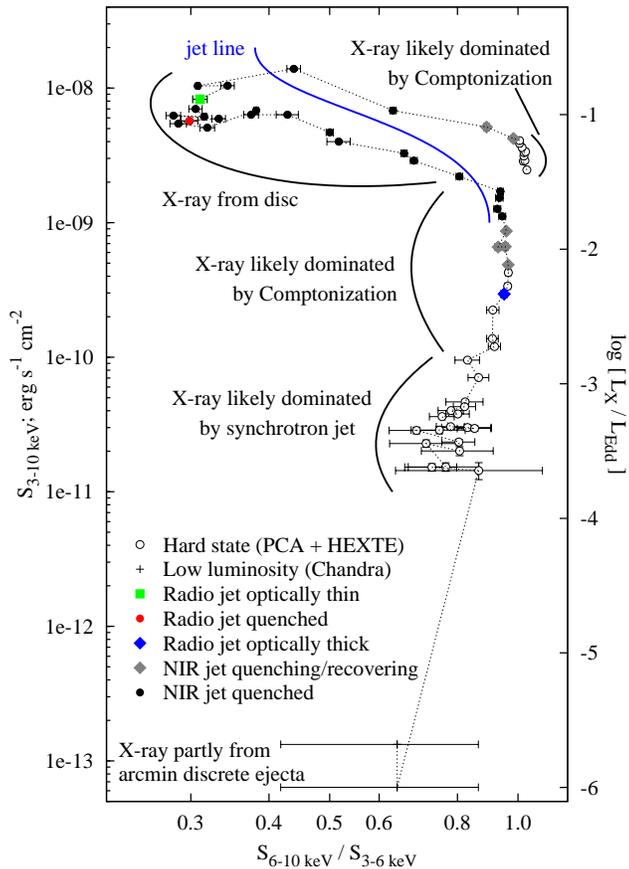}
\caption{The same X-ray HID as presented in Fig. 1, with labels marking the proposed origin of the X-ray emission in different regions of the diagram, based on the results of the multi-wavelength analysis. The proposed position of the jet line is also indicated (blue line); the steady, compact jet exists on the right hand (hard) side of the line and it is quenched on the left (soft) side.}
\end{figure}

\subsection{A revised picture for X-ray emission in BHXBs}

If the softening of the X-ray power law at low luminosities in the hard state is due to the synchrotron jet in \1550, this may also be the case for other BHXBs. Many BHXBs show this softening towards quiescence \citep{corbet06}; specifically, the majority of outbursts of sources collected in \cite{dunnet09} show this softening (using the same X-ray bands and spectral fitting methods as us) when data were obtained at low luminosities in the hard state.

The picture suggested by these data for the origin of the X-ray emission throughout the outburst of \1550 in 2000 is presented in Fig. 9. The disc, thermal Comptonization from the corona/hot inner accretion flow, the compact jet and the large-scale discrete ejecta all likely dominate the X-ray flux at different positions in the X-ray HID (the HID is the same as plotted in Fig. 1). \cite*{xueet08} found that the arcmin-scale extended X-ray jets of \1550 can produce a large fraction of the observed X-ray flux (up to 80 per cent) when the BHXB was at a low level of activity (in quiescence). The extended jet `blobs' from ejecta launched several years previously were still bright, and at times appeared brighter than the core BHXB in the 0.3--8 keV range. We include this in Fig. 9 at the lowest luminosities. It is also possible that inverse Compton emission from the base of the jet may contribute in the hard state, but if the jet is still weak initially in the hard state decline then it cannot produce the hard X-ray power law that exists before the synchrotron jet comes to dominate. In addition it seems that in this outburst, the compact jet did not return until the source was fully back in the hard state, meaning that the `jet line' \citep{fendet04} is at a harder position in the HID for the soft-to-hard transition than for the hard-to-soft transition. We indicate this jet line in blue in Fig. 9. The exact position of the jet line for the hard-to-soft transition is not well constrained for this source because no radio data were available until after this in the HID, but from compilations of other BHXBs \citep{fendet09} it likely lies close to the position indicated in Fig. 9. Only one detection of optically thin emission was made after (softer than) the jet line was crossed, probably from extended discrete ejecta and not from the compact, core jet \citep[see][for similar examples for other BHXBs]{fendet09}.

\section{Summary and conclusions}

We have re-analysed the broadband radio to X-ray evolution of the outburst of the BHXB \1550 in 2000. We show that by subtracting the decaying trend of the thermal emission in the OIR wavebands, it is possible to isolate the non-thermal excess that comes from the compact jet. We find that the quenching and recovering of the OIR jet takes several days/weeks and that on the recovery, evolution of the spectral index suggests the flow initially forms a thermal distribution of particle energies but then becomes dominated by the optically thin synchrotron component in the radiating particles.

Several lines of evidence leads us to the intriguing possibility that the synchrotron jet emission becomes dominant at X-ray wavelengths at low luminosities in the hard state. The OIR jet spectral index, the X-ray spectral index and the OIR to X-ray spectral index are all consistent with being equal and aligned ($\alpha = -0.73$), and the OIR and X-ray fluxes are linearly correlated ($L_{\rm OIR,jet} \propto L_{\rm X}^{0.98 \pm 0.08}$), strongly suggesting a single, OIR to X-ray power law from optically thin synchrotron emission declining in luminosity. In addition, the high energy cut-off or break in the hard state spectrum may shift from $\sim 100$ keV to a much lower energy, causing a steepening (softening) in the measured X-ray spectral index. In the hard state rise and initially in the hard state decline the synchrotron jet cannot produce more than a few per cent of the X-ray power law flux. As the source fades in the hard state, between $\sim 2 \times 10^{-3} $ $L_{\rm Edd}$ and $\sim 2 \times 10^{-4}$ $L_{\rm Edd}$ the most complete explanation for the evolution of the X-ray flux and spectrum is the synchrotron jet coming to dominate. The energetic state of the BHXB at these times is likely to be jet dominated \citep{fendet03}. We show that the arguments usually adopted to rule out a synchrotron jet origin to the X-ray emission cannot be successfully applied to these data. It is interesting to note that the observed jet spectral index likely remains constant ($\alpha \approx -0.73$) as the luminosity decreases by one order of magnitude, which provides an empirical constraint for jet theories. Broadband SEDs during the hard state are presented. A picture is also developed in which the origin of the X-ray emission evolves throughout this outburst and is sometimes dominated by the accretion disc, thermal Comptonization, the compact jet and discrete jets located at large distances from the binary, launched years previously.

A steepening of the X-ray power law at low luminosities in the hard state, and in quiescence is seen in many BHXBs \citep{corbet06,dunnet09}. If the origin of this steepening is the same in these systems as it is for \1550, the synchrotron jet may dominate in most BHXBs on the hard state decline below $\sim 10^{-3}$ $L_{\rm Edd}$. A corresponding change in the high energy cut-off would also be expected. Also, if the X-ray-emitting region of the jet is beamed away from the disc (i.e. if the Lorentz factor of the jet plasma at that point is high), one may expect less disc reflection of X-ray photons and a dependency on the inclination of the system. If more X-ray jets are detected for sources at different inclinations, the Lorentz factor of the compact jets may be constrained \citep[for an other method to estimate the velocity of the compact jet see][]{caseet10}.

Since this happens at low luminosities, sensitivity issues of X-ray instruments may have biased the observations in the literature to the bright hard states, when the jet does not dominate. Future sensitive X-ray instruments will have the capabilities to test this hypothesis, especially X-ray polarimeters. The optically thin synchrotron emission from the compact jet has been found to be linearly polarized and variable in the NIR \citep{shahet08,russfe08}. X-ray polarimetric studies (and variability of the polarization level and orientation) of the jet will open up a new window for studies of jet physics in accreting black holes. In addition, broadband (radio to X-ray) spectra of the isolated synchrotron jet could provide tighter constraints of the jet parameters than most modelling attempts, because normally \citep[e.g.][]{market01,market05,maitet09} the total observed SEDs of BHXBs are modelled including disc, star and other emission components.

\section*{Acknowledgments}

DMR would like to thank Tom Maccarone for a critical assessment of these results and valuable discussions, and Rob Fender also for helpful discussions. We thank the referee for a swift, thorough response. These results are based on data taken with the YALO telescope (currently operated by the SMARTS consortium), NASA's RXTE and Chandra X-ray missions and the Australia Telescope Compact Array. DMR and SM acknowledge support from a Netherlands Organization for Scientific Research (NWO) Veni and Vidi Fellowship, respectively. RJHD acknowledges support from the Alexander von Humboldt Foundation. This research was supported by the DFG cluster of excellence `Origin and Structure of the Universe' (www.universe-cluster.de).

\bsp

\label{lastpage}


\begin{thebibliography}{99}
\bibitem[\protect\citeauthoryear{Aref'ev et al.}{2004}]{arefet04}Aref'ev V. A., Revnivtsev M. G., Lutovinov A. A., Sunyaev R. A., 2004, AstL, 30, 669
\bibitem[\protect\citeauthoryear{Belloni}{2009}]{bell09}Belloni T. M., 2010, in `The Jet Paradigm - From Microquasars to Quasars', ed. T. Belloni, Lect. Notes Phys. 794, Springer-Verlag Berlin, Heidelberg, p. 53 (arXiv:0909.2474)
\bibitem[\protect\citeauthoryear{Belloni et al.}{2002}]{bellet02}Belloni T., Colombo A. P., Homan J., Campana S., van der Klis M., 2002, A\&A, 390, 199
\bibitem[\protect\citeauthoryear{Blandford \& Konigl}{1979}]{blanko79}Blandford R. D., Konigl A., 1979, ApJ, 232, 34
\bibitem[\protect\citeauthoryear{Buxton \& Bailyn}{2004}]{buxtba04}Buxton, M. M., Bailyn, C. D. 2004, ApJ, 615, 880
\bibitem[\protect\citeauthoryear{Cabanac et al.}{2009}]{cabaet09}Cabanac C., Fender R. P., Dunn R. J. H., K\"ording E. G., 2009, MNRAS, 396, 1415
\bibitem[\protect\citeauthoryear{Cardelli, Clayton \& Mathis}{Cardelli et al.}{1989}]{cardet89}Cardelli J. A., Clayton G. C., Mathis J. S., 1989, ApJ, 345, 245
\bibitem[\protect\citeauthoryear{Casella et al.}{2010}]{caseet10}Casella P., et al., 2010, MNRAS Letters, in press (arXiv:1002.1233)
\bibitem[\protect\citeauthoryear{Charles \& Coe}{2006}]{charco06}Charles P. A., Coe M. J., 2006, in Compact Stellar X-Ray Sources, eds. Lewin W. H. G., van der Klis M., Cambridge University Press, p. 215
\bibitem[\protect\citeauthoryear{Corbel et al.}{2000}]{corbet00}Corbel S., Fender R. P., Tzioumis A.K., Nowak M., McIntyre V., Durouchoux P., Sood R., 2000, A\&A, 359, 251
\bibitem[\protect\citeauthoryear{Corbel et al.}{2001}]{corbet01}Corbel S., et al., 2001, ApJ, 554, 43
\bibitem[\protect\citeauthoryear{Corbel \& Fender}{2002}]{corbfe02}Corbel S., Fender R. P., 2002, ApJ, 573, L35
\bibitem[\protect\citeauthoryear{Corbel et al.}{2002}]{corbet02}Corbel S., Fender R. P., Tzioumis A. K., Tomsick J. A., Orosz J. A., Miller J. M., Wijnands R., Kaaret P., 2002, Sci, 298, 196
\bibitem[\protect\citeauthoryear{Corbel et al.}{2003}]{corbet03}Corbel S., Nowak M. A., Fender R. P., Tzioumis A. K., Markoff, S. 2003, A\&A, 400, 1007
\bibitem[\protect\citeauthoryear{Corbel, Tomsick \& Kaaret}{Corbel et al.}{2006}]{corbet06}Corbel S., Tomsick J. A., Kaaret P., 2006, ApJ, 636, 971
\bibitem[\protect\citeauthoryear{Coriat et al.}{2009}]{coriet09}Coriat M., Corbel S., Buxton M. M., Bailyn C. D., Tomsick J. A., K\"ording E., Kalemci E., 2009, MNRAS, 400, 123
\bibitem[\protect\citeauthoryear{Done \& Gierli\'nski}{2003}]{donegi03}Done C., Gierli\'nski M., 2003, MNRAS, 342, 1041
\bibitem[\protect\citeauthoryear{Done, Gierli\'nski \& Kubota}{Done et al.}{2007}]{doneet07}Done C., Gierli\'nski M., Kubota A., 2007, A\&ARv, 15, 1
\bibitem[\protect\citeauthoryear{Dunn et al.}{2008}]{dunnet08}Dunn R. J. H., Fender R. P., K\"ording E. G., Cabanac C., Belloni T., 2008, MNRAS, 387, 545
\bibitem[\protect\citeauthoryear{Dunn et al.}{2009}]{dunnet09}Dunn R. J. H., Fender R. P., K\"ording E. G., Belloni T., Cabanac C., 2009, MNRAS, in press (arXiv:0912.0142)
\bibitem[\protect\citeauthoryear{Falcke \& Biermann}{1995}]{falcbi95}Falcke H., Biermann P. L., 1995, A\&A, 293, 665 
\bibitem[\protect\citeauthoryear{Falcke \& Biermann}{1996}]{falcbi96}Falcke H., Biermann P. L., 1996, A\&A, 308, 321
\bibitem[\protect\citeauthoryear{Falcke, K\"ording \& Markoff}{Falcke et al.}{2004}]{falcet04}Falcke H., K\"ording E., Markoff S., 2004, A\&A, 414, 895
\bibitem[\protect\citeauthoryear{Fender}{2001}]{fend01}Fender R. P., 2001, MNRAS, 322, 31
\bibitem[\protect\citeauthoryear{Fender}{2006}]{fend06}Fender R. P., 2006, in Compact Stellar X-Ray Sources, eds. Lewin W. H. G., van der Klis M., Cambridge University Press, p. 381
\bibitem[\protect\citeauthoryear{Fender, Gallo \& Jonker}{Fender et al.}{2003}]{fendet03}Fender R. P., Gallo E., Jonker P. G., 2003, MNRAS, 343, L99
\bibitem[\protect\citeauthoryear{Fender, Belloni \& Gallo}{Fender et al.}{2004}]{fendet04}Fender, R. P., Belloni, T. M., Gallo, E. 2004, MNRAS, 355, 1105
\bibitem[\protect\citeauthoryear{Fender et al.}{2009}]{fendet09}Fender R. P., Homan J., Belloni T. M., 2009, MNRAS, 396, 1370
\bibitem[\protect\citeauthoryear{Gallo, Fender \& Pooley}{Gallo et al.}{2003}]{gallet03}Gallo, E., Fender, R. P., Pooley, G. G. 2003, MNRAS, 344, 60
\bibitem[\protect\citeauthoryear{Gallo et al.}{2005}]{gallet05}Gallo E., Fender R. P., Kaiser C., Russell, D. M., Morganti R., Oosterloo T., Heinz S., 2005, Nat, 436, 819
\bibitem[\protect\citeauthoryear{Gallo et al.}{2006}]{gallet06}Gallo E., et al., 2006, MNRAS, 370, 1351
\bibitem[\protect\citeauthoryear{Gierli\'nski \& Newton}{2006}]{gierne06}Gierli\'nski M., Newton J., 2006, MNRAS, 370, 837
\bibitem[\protect\citeauthoryear{Gilfanov}{2009}]{gilf09}Gilfanov M., 2009, in `The Jet Paradigm - From Microquasars to Quasars', ed. T. Belloni, Lect. Notes Phys. 794, Springer-Verlag Berlin, Heidelberg, p. 17 (arXiv:0909.2567)
\bibitem[\protect\citeauthoryear{Grove et al.}{1998}]{grovet98}Grove J. E., Johnson W. N., Kroeger R. A., McNaron-Brown K., Skibo J. G., Phlips B. F., 1998, ApJ, 500, 899
\bibitem[\protect\citeauthoryear{G\"ultekin et al.}{2009}]{gultet09}G\"ultekin K., Cackett E. M., Miller J. M., Di Matteo T., Markoff S., Richstone D. O., 2009, ApJ, 706, 404
\bibitem[\protect\citeauthoryear{Heinz}{2004}]{hein04}Heinz S., 2004, MNRAS, 355, 835
\bibitem[\protect\citeauthoryear{Heinz \& Sunyaev}{2003}]{heinsu03}Heinz S., Sunyaev R. A., 2003, MNRAS, 343, L59
\bibitem[\protect\citeauthoryear{Homan et al.}{2005}]{homaet05}Homan J., Buxton M., Markoff S., Bailyn C. D., Nespoli E., Belloni T., 2005, ApJ, 624, 295
\bibitem[\protect\citeauthoryear{Hynes}{2005}]{hyne05}Hynes R. I., 2005, ApJ, 623, 1026
\bibitem[\protect\citeauthoryear{Hynes et al.}{2003}]{hyneet03}Hynes R. I., et al., 2003, MNRAS, 345, 292
\bibitem[\protect\citeauthoryear{Jain et al.}{2001}]{jainet01}Jain R. K., Bailyn C. D., Orosz J. A., McClintock J. E., Remillard R. A., 2001, ApJ, 554, L181
\bibitem[\protect\citeauthoryear{Joinet, Kalemci \& Senziani}{Joinet et al.}{2008}]{joinet08}Joinet A., Kalemci E., Senziani F., 2008, ApJ, 679, 655
\bibitem[\protect\citeauthoryear{Kaaret et al.}{2003}]{kaaret03}Kaaret P., Corbel S., Tomsick J. A., Fender R., Miller J. M., Orosz J. A., Tzioumis A. K., Wijnands R., 2003, ApJ, 582, 945
\bibitem[\protect\citeauthoryear{Kaiser}{2005}]{kais05}Kaiser C. R., 2005, MNRAS, 360, 176
\bibitem[\protect\citeauthoryear{Kalemci et al.}{2001}]{kaleet01}Kalemci E., Tomsick J. A., Rothschild R. E., Pottschmidt K., Kaaret P., 2001, ApJ, 563, 239
\bibitem[\protect\citeauthoryear{Kanbach et al.}{2001}]{kanbet01}Kanbach G., Straubmeier C., Spruit H. C., Belloni T., 2001, Nat, 414, 180
\bibitem[\protect\citeauthoryear{K\"ording, Falcke \& Corbel}{K\"ording et al.}{2006a}]{kordet06a}K\"ording E., Falcke H., Corbel S., 2006a, A\&A, 456, 439
\bibitem[\protect\citeauthoryear{K\"ording, Fender \& Migliari}{K\"ording et al.}{2006b}]{kordet06b}K\"ording E., Fender R. P., Migliari S., 2006b, MNRAS, 369, 1451
\bibitem[\protect\citeauthoryear{K\"ording et al.}{2008}]{kordet08}K\"ording E., Rupen M., Knigge C., Fender R., Dhawan V., Templeton M., Muxlow T., 2008, Sci, 320, 1318
\bibitem[\protect\citeauthoryear{Maccarone}{2005}]{macc05}Maccarone T. J., 2005, MNRAS, 360, L68
\bibitem[\protect\citeauthoryear{Maccarone \& Coppi}{2003}]{maccco03}Maccarone T. J., Coppi P. S., 2003, MNRAS, 338, 189
\bibitem[\protect\citeauthoryear{Maitra et al.}{2009}]{maitet09}Maitra D., Markoff S., Brocksopp C., Noble M., Nowak M., Wilms J., 2009, MNRAS, 398, 1638
\bibitem[\protect\citeauthoryear{Malzac, Merloni \& Fabian}{Malzac et al.}{2004}]{malzet04}Malzac J., Merloni A., Fabian A. C., 2004, MNRAS, 351, 253
\bibitem[\protect\citeauthoryear{Markoff, Falcke \& Fender}{Markoff et al.}{2001}]{market01}Markoff S., Falcke H., Fender R., 2001, A\&A, 372, L25
\bibitem[\protect\citeauthoryear{Markoff et al.}{2003}]{market03}Markoff, S., Nowak, M., Corbel, S., Fender, R., Falcke, H. 2003, A\&A, 397, 645
\bibitem[\protect\citeauthoryear{Markoff, Nowak \& Wilms}{Markoff et al.}{2005}]{market05}Markoff S., Nowak M. A., Wilms J., 2005, ApJ, 635, 1203
\bibitem[\protect\citeauthoryear{Merloni, Heinz \& di Matteo}{Merloni et al.}{2003}]{merlet03}Merloni A., Heinz S., di Matteo T., 2003, MNRAS, 345, 1057
\bibitem[\protect\citeauthoryear{Migliari \& Fender}{2006}]{miglfe06}Migliari S., Fender R. P., 2006, MNRAS, 366, 79
\bibitem[\protect\citeauthoryear{Migliari et al.}{2006}]{miglet06}Migliari S., Tomsick J. A., Maccarone T. J., Gallo E., Fender R. P., Nelemans G., Russell D. M., 2006, ApJ, 643, L41
\bibitem[\protect\citeauthoryear{Migliari et al.}{2007}]{miglet07}Migliari S., et al., 2007, ApJ, 670, 610
\bibitem[\protect\citeauthoryear{Miller et al.}{2001}]{millet01}Miller J. M., et al., 2001, ApJ, 563, 928
\bibitem[\protect\citeauthoryear{Mirabel et al.}{1992}]{miraet92}Mirabel I. F., Rodr\'{i}guez L. F., Cordier B., Paul J., Lebrun F., 1992, Nat, 358, 215
\bibitem[\protect\citeauthoryear{Miyakawa et al.}{2008}]{miyaet08}Miyakawa T., Yamaoka K., Homan J., Saito K., Dotani T., Yoshida A., Inoue H., 2008, PASJ, 60, 637
\bibitem[\protect\citeauthoryear{Motta, Belloni \& Homan}{Motta et al.}{2009}]{mottet09}Motta S., Belloni T., Homan J., 2009, MNRAS, 400, 1603
\bibitem[\protect\citeauthoryear{Narayan \& Yi}{1995}]{narayi95}Narayan R., Yi I, 1995, ApJ, 452, 710
\bibitem[\protect\citeauthoryear{Nowak et al.}{2005}]{nowaet05}Nowak M. A., Wilms J., Heinz S., Pooley G., Pottschmidt K., Corbel S., 2005, ApJ, 626, 1006
\bibitem[\protect\citeauthoryear{Orosz et al.}{2002}]{oroset02}Orosz J. A., et al., 2002, ApJ, 568, 845
\bibitem[\protect\citeauthoryear{Reilly et al.}{2001}]{reilet01}Reilly K. T., et al., 2001, ApJ, 561, L183
\bibitem[\protect\citeauthoryear{Rodriguez, Corbel \& Tomsick}{Rodriguez et al.}{2003}]{rodret03}Rodriguez J., Corbel S., Tomsick J. A., 2003, ApJ, 595, 1032
\bibitem[\protect\citeauthoryear{Russell et al.}{2006}]{russet06}Russell D. M., Fender R. P., Hynes R. I., Brocksopp C., Homan J., Jonker P. G., Buxton M. M., 2006, MNRAS, 371, 1334
\bibitem[\protect\citeauthoryear{Russell et al.}{2007}]{russet07}Russell D. M., Maccarone T. J., K\"ording E. G., Homan J., 2007, MNRAS, 379, 1401
\bibitem[\protect\citeauthoryear{Russell \& Fender}{2008}]{russfe08}Russell D. M., Fender R. P., 2008, MNRAS, 387, 713
\bibitem[\protect\citeauthoryear{Russell et al.}{2008}]{russet08}Russell D. M., Maitra D., Fender R. P., Lewis F., 2008, in VII Microquasar Workshop: Microquasars and Beyond (arXiv:0811.2919)
\bibitem[\protect\citeauthoryear{Shahbaz et al.}{2008}]{shahet08}Shahbaz T., Fender R. P., Watson C. A., O'Brien K., 2008, ApJ, 672, 510
\bibitem[\protect\citeauthoryear{Sunyaev \& Titarchuk}{1980}]{sunyti80}Sunyaev R. A., Titarchuk L. G., 1980, A\&A, 86, 121
\bibitem[\protect\citeauthoryear{Tomsick, Corbel \& Kaaret}{Tomsick et al.}{2001}]{tomset01}Tomsick J. A., Corbel S., Kaaret P., 2001, ApJ, 563, 229
\bibitem[\protect\citeauthoryear{Tomsick et al.}{2003}]{tomset03}Tomsick J. A., Corbel S., Fender R., Miller J. M., Orosz J. A., Tzioumis T., Wijnands R., Kaaret P., 2003, ApJ, 582, 933
\bibitem[\protect\citeauthoryear{Xue, Wu \& Cui}{Xue et al.}{2008}]{xueet08}Xue Y., Wu X.-B., Cui W., 2008, MNRAS, 384, 440
\bibitem[\protect\citeauthoryear{Yuan et al.}{2007}]{yuanet07}Yuan F., Zdziarski A. A., Xue Y., Wu X.-B., 2007, ApJ, 659, 541
\bibitem[\protect\citeauthoryear{Zdziarski et al.}{2003}]{zdziet03}Zdziarski A. A., Lubi\'nski P., Gilfanov M., Revnivtsev M., 2003, MNRAS, 342, 355


\end{thebibliography}
\end{document}